%% file: main.tex
\documentclass[conference]{IEEEtran}
\usepackage{cite}
\usepackage{amsmath,amssymb,amsfonts}
\usepackage{algorithmic}
\usepackage{graphicx}
\usepackage{textcomp}
\usepackage{xcolor}
\usepackage{makecell}
\usepackage{colortbl} 
\usepackage{comment}
\usepackage{flushend}
\usepackage{multirow}\usepackage{multirow}
\usepackage{rotating}
\usepackage{hyperref}

\begin{document}

\title{Analyzing Social Media Engagement of Computer Science Conferences}

\author{
\IEEEauthorblockN{Rey Ortiz}
\IEEEauthorblockA{\textit{Department of Computer Science} \\
\textit{California Polytechnic State University}\\
San Luis Obispo, USA \\
dortiz11@calpoly.edu}
    \and

\IEEEauthorblockN{Sharif Ahmed}
\IEEEauthorblockA{\textit{Computer Science Department} \\
\textit{Boise State University}\\
Boise, USA \\
sharifahmed@u.boisestate.edu}
    \and

\IEEEauthorblockN{Priscilla Salas}
\IEEEauthorblockA{\textit{Department of Computer Science} \\
\textit{Allan Hancock College}\\
Santa Maria, USA \\
priscill.salas@my.hancockcollege.edu}
    \and
    
\IEEEauthorblockN{Nasir U. Eisty}
\IEEEauthorblockA{\centerline{\textit{Computer Science Department}} \\
\textit{Boise State University}\\
Boise, Idaho, USA \\
nasireisty@boisestate.edu}
   }


\maketitle
\begin{abstract}
\input{abstract}
\end{abstract}


\section{ Introduction}
\label{sec_introduction}
\input{sec_introduction}

\section{ Background}
\label{sec_background}
\input{sec_background}

\section{ Methodology}
\label{sec_methodology}

\input{sec_methodology}

\section{ Results and Discussion}
\label{sec_result_and_discussion}
\input{sec_result_and_discussion}

\section{ Threats to Validity}
\label{sec_threats}
\input{sec_threats}

\section{ Conclusion}
\label{sec_conclusion}

\input{sec_conclusion}

\section*{Data Availability}
\label{sec_data}
The data and artifacts of this paper can be accessed  at -\\
 \textcolor{blue}{\href{https://figshare.com/s/c9fbb8513d519384efe3}{https://figshare.com/s/c9fbb8513d519384efe3}}

\bibliographystyle{IEEEtran}
\bibliography{references}
\end{document}

%% file: abstract.tex
\textit{Context:} 
X, formerly known as Twitter, is one of the largest social media platforms and has been widely used for communication during research conferences. While previous studies have examined how users engage with X during these events, limited research has focused on analyzing the content posted by computer science conferences. 
\textit{Objective:} 
This study investigates how conferences from different areas of computer science perform on social media by analyzing their activity, follower engagement, and the content posted on X.
\textit{Method:} 
We collect posts from 22 computer science conferences and conduct statistical experiments to identify variations in content. Additionally, we perform a manual analysis of the top five posts for each engagement metric.
\textit{Results:} 
Our findings indicate statistically significant differences in category, sentiment, and post length across computer science conference posts. Among all engagement metrics, likes were the most common way users interacted with conference content.
\textit{Conclusion:} 
This study provides insights into the social media presence of computer science conferences, highlighting key differences in content, sentiment, and engagement patterns across different venues.

%% file: sec_introduction.tex
X\footnote{https://x.com} (formerly Twitter) serves as a backchannel for conferences and has been recognized as an effective platform for fostering a ``more participatory conference culture'' by enhancing communication and engagement~\cite{schmitt2017computer}~\cite{ross2011enabled}. Prior research has examined user behavior on Twitter during conferences.

Wen et al.~\cite{wen2014groups} explored how different user groups (\textit{Junior Researcher, Senior Researcher, Faculty, Industry}, and \textit{Organization}) interacted on Twitter during academic conferences. Their study, which analyzed four computer science conferences in 2012, found that interactions were primarily among peers with similar levels of experience within the community. In three of the four conferences, newcomer participation was limited, and the attention they received was even lower. This finding challenges the hypothesis proposed by Reinhardt et al.~\cite{reinhardt2009people}, which suggested that Twitter could be beneficial for newcomers to engage within the academic community.

Parra et al.~\cite{parra2016twitter} analyzed 16 computer science conferences over a four-year period (2009–2013) to examine how Twitter was used for communication during scientific events over time. Their study also aimed to identify key characteristics, such as "topical trends and sentiment," that could potentially predict whether a user would return to a scientific event.

While previous studies have examined user behavior during conferences and analyzed tweets using specific event hashtags (e.g., ``\#www'' for the WWW conference)~\cite{wen2015information}~\cite{parra2016twitter}, there remains a gap in understanding the broader social media presence of computer science conferences. This paper aims to address that gap by investigating the role of follower count in engagement metrics, analyzing how conference content varies across events, and exploring whether the content itself influences user interactions with conference posts. Our study involves scraping all posts from 22 computer science conferences over 14 years, thus making it, to our knowledge, the first study to analyze such an extensive time frame and dataset within this domain.

Our interest in exploring the relationship between follower count and influence led us to investigate whether accounts with more followers tend to be more popular and impactful. On social media, an account’s follower count is often indicative of its reach and influence. This is particularly evident among social media influencers, whose large followings enable them to significantly shape their audience’s decision-making~\cite{hudders2021commercialization}. As a result, influencers are frequently sought after by advertisers for brand endorsements. 
Additionally, prior research suggests that influence occurs when information is perceived and acted upon, primarily through communicated content rather than personal attributes such as family connections or celebrity status~\cite{anger2011measuring}. When user A follows user B and engages with their posts, whether through retweets or comments, they demonstrate being influenced by that content.  
This idea led us to formulate our first research question:  

\textbf{RQ1: Does a computer science conference account’s follower count influence audience engagement with its content?}

Understanding this relationship between follower count and engagement metrics is essential, as a higher follower count is often assumed to translate into greater engagement. Still, the extent to which this holds for academic conferences remains unclear. So, we formulated the following research question:  

\textbf{RQ2: Is there a correlation between the follower count and engagement metrics for computer science conference accounts on X?}

A previous study explored whether user feedback on X varied across different countries based on tweet characteristics such as length, content, and sentiment~\cite{tabbassum2023towards}. Inspired by this, we sought to investigate these characteristics in the context of computer science conferences to understand how their social media presence differs. Identifying variations in content, sentiment, and post length can provide insights into how conferences engage their audiences and whether certain factors contribute to higher engagement.  
To address this, we formulated the following research questions: 

\textbf{RQ3.1: Does the content of posts differ significantly across selected conferences?}

\textbf{RQ3.2: Are there significant differences in sentiment scores across selected conferences?}

\textbf{RQ3.3: Does the length of posts vary significantly across selected conferences?}

Driven by the need to understand how the computer science community interacts with conference-related content on social media, we aim to explore user engagement patterns. Analyzing these interactions can provide valuable insights into which types of content resonate most with the audience and how engagement varies across different conferences. This leads us to the final research question:  

\textbf{RQ4: How do users on X engage with content posted by computer science conference accounts?}

In summary, our study aims to bridge the gap in understanding how computer science conferences utilize social media and how their content resonates with the community. By analyzing follower influence, engagement metrics, content differences, and user interactions, we seek to uncover key factors that shape the online presence of these conferences. The insights gained from this research can help conferences optimize their social media strategies, enhance audience engagement, and foster a more interactive and participatory research community.

%% file: sec_background.tex
This section outlines key metrics associated with an X (formerly Twitter) account and the engagement metrics that reflect user interactions with posts. These metrics provide a comprehensive view of how users interact with content, offering insights into the impact and visibility of posts from computer science conference accounts.

\begin{itemize}
    \item \textbf{Followers:} The number of users who follow a particular X account.
    \item \textbf{Following:} The number of users that a particular X account follows.
    \item \textbf{Likes:} A metric that indicates positive sentiment toward a post on X.
    \item \textbf{Replies:} The number of direct responses a post on X receives, reflecting engagement and discussion.
    \item \textbf{Reposts:} Similar to likes in expressing approval, a repost occurs when an X user shares someone else's post with their followers on X.
    \item \textbf{Shares:} The number of times an X post has been reposted with an added comment, contributing to further discussion and context.
\end{itemize}

%% file: sec_methodology.tex
This section describes our methodology for data collection and the approach we utilized to analyze the statistical evidence.

\subsection{Data Collection}

Initially, we explored using the X API to collect posts from various computer science conference accounts. However, during the early stages of this process, we encountered restrictions that prevented us from scraping posts, as X has recently imposed stricter limitations on data access. To overcome this issue, we sought alternative solutions and identified a tool that leverages Nitter\footnote{https://github.com/bocchilorenzo/ntscraper} instances, allowing us to collect data without relying on the X API. Unlike the X API, which enforces limits on the number of posts that can be retrieved, this tool provided unrestricted access to historical posts.  

For our study, we focused on collecting posts from the top-ranked conferences across different fields of computer science. To determine these conferences, we used a website (https://csconferences.org), which categorizes conferences based on different research areas and provides acceptance rates over a five-year period. We manually selected the top-ranked conference in each field by identifying those with the lowest acceptance rates. Once the top conference for each field was selected, we used the Nitter tool to scrape posts by specifying the X handle of each conference. Our data collection spanned from January 1, 2010, to February 10, 2024, the date we initiated the scraping process. This resulted in a dataset of 10,187 posts from 22 conferences.  

To gather information on follower counts and the number of accounts each conference follows, we manually visited each conference's X profile and recorded this data. To ensure consistency, all data was collected on the same day and within the same time window, minimizing potential discrepancies.

\subsection{Sample Creation}

Given the large volume of 10,187 posts in our dataset, manually inspecting each one would be highly time-consuming. To make the analysis feasible, we constructed a representative sample using a 95\% confidence interval.  
To create this sample, we first considered the total number of posts collected from all conference X accounts, denoted as N1 = 10,187. We then calculated 95\% confidence interval of N1, which yielded N2 = 344, representing the target sample size.  
Next, we determined the proportion of posts contributed by each individual conference by dividing the total number of posts from that conference by N1. We then multiplied this proportion by N2 to allocate an appropriate number of posts from each conference into the final sample.  
Through this approach, we arrived at a final sample size of 352 posts, ensuring a balanced and statistically representative subset for our manual analysis.

\subsection{Manual Annotation}

Our more manageable sample dataset of 352 posts allowed two of the authors to annotate and interpret each post manually. This manual approach was chosen over automated classification methods, as prior research has shown that automated sentiment analysis can often produce unreliable or less accurate results than human evaluation~\cite{barbera2021automated, boukes2020s, mandliautomated}.  
This manual sentiment classification enabled a more reliable and nuanced understanding of how users engage with conference posts.
For sentiment evaluation, we adopted a well-established coding scheme used in previous studies~\cite{nugroho2021project}. The classification criteria are as follows:  

\begin{itemize}
  \item \textbf{Positive:} Posts that convey a positive sentiment or include positive words (e.g., nice, working example, thanks a lot).  
  \item \textbf{Negative:} Posts that express a negative sentiment or contain negative words (e.g., error, bug, not working).  
  \item \textbf{Neutral:} Posts that are emotionally neutral, either balancing both positive and negative expressions or lacking any explicitly biased words (i.e., neither strongly positive nor negative).  
\end{itemize}

\input{tbl/category_tbl}

To better understand the type of content posted by computer science conferences on X, we categorized posts from our sample dataset of 352 tweets. The initial classification framework was adapted from a previous study~\cite{sharma2015s}, serving as a foundation for our categorization process. These predefined categories, along with their descriptions, are presented in Table~\ref{category_tbl}.  
During the manual annotation process, we expanded the category list by individually analyzing the content of posts in our sample. Following the approach Guzman et al.~\cite{guzman2016needle} used, we defined new categories with precise descriptions and relevant examples. We merged similar categories and refined their definitions accordingly before finalizing 
Table~\ref{category_tbl}.  

To ensure a systematic classification, we adhered to the \textit{Content Analysis Guidebook} by Neuendorf~\cite{neuendorf2017content}, which emphasizes that categories should be exhaustive, mutually exclusive, and at an appropriate level of measurement. In line with this guideline, each post was assigned to only one category. Additionally, to account for tweets that did not fit any defined category, we included an ``Other'' category as recommended by the guide.  

To further validate the manual annotation process for both sentiment and content classification, we measured \textit{inter-annotator agreement} using the \textit{Kappa calculator}\footnote{http://justusrandolph.net/kappa/} on the 352 sampled tweets~\cite{viera2005understanding}. The Kappa score for sentiment classification was 0.61, while for content categorization, it was 0.67, both indicating substantial agreement within acceptable reliability ranges. Any disagreements between annotators were resolved through face-to-face discussions before finalizing the annotations~\cite{ahmed2024understanding}.  
This categorization process ensures a structured and reliable classification of conference posts, providing valuable insights into how different conferences engage with their audiences on X.

\subsection{Experiments}

\subsubsection{\textbf{RQ1: Followers Count Influence}}

X has various metrics that can be used to determine a user’s account performance, such as the number of followers, the number of accounts a user is following, posts, and reposts~\cite{anger2011measuring}. These metrics can be used in conjunction with the following ratios: follower/following ratio,  retweet ratio, and interactor ratio~\cite{riquelme2016measuring}~\cite{anger2011measuring}.

The \textit{Follower/Following ratio (FFr)} is a comparison of the number of followers by the number of users that a particular account is following. If the resulting value is one-to-one, this is an indication that the account is a mass follower. In contrast, an account with a higher value is an indication that other users are interested in the content the account is posting. The formula for this is the following:

\[FFr(i) = \frac{\textnormal{\# of followers}}{\textnormal{\# of users i is following}}\]

\vspace{1mm}
The \textit{Repost and Mention ratio (RMr)} is an indication of how many of i’s (conference account) posts imply a reaction from the audience. The formula for this is as follows:

\[RMr(i) = \frac{\textnormal{\# of posts of i reposted + \# posts of i replied}}{\textnormal{\# of posts of i}}\]
\vspace{1mm}

The \textit{Interactor ratio (Ir)} is used to measure the number of users that interact with i. The formula for this is as follows:

\[Ir(i) = \frac{\textnormal{\# of users who have reposted i posts}}{\textnormal{\# of followers of i}}\]
\vspace{1mm}

To evaluate the influence and engagement figures, we ranked the conferences by the number of followers, a metric used to gauge success on X. Table~\ref{tbl_conf_ratios} shows the conference X handle, the category of the conference, the number of followers and followees, and the results from the ratios that determine an account's performance.

The FFr alone cannot indicate if an account is influential since it must be interpreted together with the RMr and Ir. It does however indicate if other users find a X account interesting. The RMr indicates how many posts imply a reaction while the Ir measures number of user that interact with the conference's post(s). For an account to be influential it will have a high FFr ratio along with the RMr and Ir since if an account has more followers there is a higher likelihood that they are going to engage with the content.

\subsubsection{\textbf{RQ2: Followers Count vs Engagement Metrics}}

To answer this RQ, we consider the total number of followers of a conference's X account and the following X engagement metrics: likes, reposts, shares, and replies.  We obtained the total number of followers from all of the conferences as well as the total of each of the engagement metrics. Subsequently, we performed a Pearson correlation check with the number of followers and the engagement metrics.

\subsubsection{\textbf{RQ3.1: Content Diversity and RQ3.2: Sentiment Diversity}}

We used the manually annotated categorical post data for the Chi-Square test to identify differences in content across conferences. If the Chi-Square test produced a p-value$<$0.05, we then applied the Bonferroni Adjustment Post Hoc test to further analyze the significance of these differences. We then determined which conferences showed statistically significant differences based on the threshold values of 0.05, 0.01, and 0.001 from the Bonferroni Adjustment Post Hoc test results.

\subsubsection{\textbf{RQ3.3: Post length diversity}}

We first cleaned our dataset to ensure that post lengths were not skewed by unnecessary elements. We removed whitespace, emojis, and ampersands, as some users tend to overuse them. Then, we calculated the length of each post and applied the Kruskal-Wallis test to detect statistical differences between selected conferences. If the p-value$<$0.05, we performed the Dunn Post Hoc test to determine which specific conferences were significantly different from each other. The Dunn test provided p-values for each conference pair, and we identified statistically significant differences based on the threshold values of 0.05, 0.01, and 0.001.  
We performed this test on both our sample dataset (352 posts) and the complete dataset (10,187 posts) to ensure consistency and robustness in our analysis.

\subsubsection{\textbf{RQ4: User Engagement}}

To understand how X users respond to posts, we looked at how a user can respond to a post. Users can respond to a post by commenting, reposting, sharing, and liking. To analyze these metrics, we applied basic descriptive statistical measurements such as mean and median at each of the conferences' collective posts. 
Lastly, we opt for a manual analysis following existing research~\cite{ahmed2024decade}. 
Initially, we composed a list of the top 5 posts for each engagement metric. 
Then, we manually analyzed the top 5 tweets from each metric to identify and reach possible conclusions as to why a tweet ranked in the top 5 for the particular metric.

%% file: tbl/category_tbl.tex
\begin{table*}[hbtp]
\caption{Categories of Posts in Computer Science Conferences}
\renewcommand{\arraystretch}{1.3}
\centering
\label{category_tbl}
\resizebox{2\columnwidth}{!}{%
\begin{tabular}{|p{0.2\linewidth}|p{0.68\linewidth}|p{0.12\linewidth}|}
\hline
\textbf{Category}              & \textbf{Description}                                                                                                & \textbf{Source} \\ \hline
Article and Multimedia Sharing & Tweets sharing articles, blogs, tutorials, or videos related to computer science                                    & Sharma et al.~\cite{sharma2015s}   \\ \hline
Awards                         & Posts on X sharing recognition, awards nominations, presenting awards , announce winners (ie, best paper, etc)           & This work        \\ \hline
Career                         & Tweets about job openings and volunteering opportunities                                                            & Sharma et al.~\cite{sharma2015s}   \\ \hline
Community Events               & Tweets about conferences, coding events, workshops, etc                                                             & Sharma et al.~\cite{sharma2015s}   \\ \hline
Crowdsourcing Requests          & Tweets requesting users to contribute to open source projects, surveys, petitions, etc. related to computer science & Sharma et al.~\cite{sharma2015s}   \\ \hline
Resources                      & Posts on X sharing information pertaining to grants, assistance and access to the conference                            & This work        \\ \hline
Satires                        & Tweets sharing jokes and funny quotes                                                                               & Sharma et al.~\cite{sharma2015s}   \\ \hline
Submissions                    & Posts on X announcing dates for submissions and/or extensions for papers, awards, demos, etc                            & This work        \\ \hline
Others                         & All other tweets which do not fall into one of the above categories                                                 & Sharma et al.~\cite{sharma2015s}   \\ \hline

\end{tabular}
}
\end{table*}

%% file: sec_result_and_discussion.tex
This section reports and discusses the results of our study.

\input{tbl/tbl_conf_ratios}
\subsection{\textbf{RQ1: Does a computer science conference account’s follower count influence audience engagement with its content?}}

The conferences leading the list of the most followers are conferences from the categories ICML (machine learning), CVPR (computer vision), and SIGGRAPH (computer graphics), respectively, as shown in Table~\ref{tbl_conf_ratios}. While the Follower/Following ratio (FFr) is a performance indicator, a higher value indicates that more people are interested in the user's posts. The ICML Conference was the leading conference for the FFr with 4632.2. For each account they follow, they have 4632.2 followers. This value alone can suggest that the more followers an account has because of its influence since other users follow what they are interested in. This alone can lead to misinterpretation; hence, we must look at them with respect to the other interaction ratios.

Regarding the RMr ranking from the sample dataset, the ICML conference was ranked 19 and had a ranking of 16 from the complete dataset of posts. The CVPR and SIGGRAPH conferences showed similar results, with the CVPR conference ranked 16 in the sample dataset and 20 in the complete dataset. SIGGRAPH had the lowest ranking of 22 in the sample dataset and 20 overall in the complete dataset. The PPoPP (Programming Languages) conference had the top rank for RMi in the sample dataset, and the UIST (Human-Computer Interaction) conference had the top rank for the RMi for the complete dataset. The PPoPP conference was second to last in terms of followers.

The PPoPP conference was the top rank in the Interactor ratio, Ir, for the sample dataset, while the UIST conference had the top rank for the Ir of the complete dataset. The leading conferences in terms of followers failed to make it to the top 10 for both datasets for the Ir.

 From this analysis, we have observed that having a higher number of followers does not indicate that a user will be influenced to engage with the content. We observed that conferences with fewer followers had users engage with their content compared to the top conferences with more followers.

\input{tbl/tbl_corr_conf}
\subsection{\textbf{RQ2: Is there a correlation between the follower count and engagement metrics for computer science conference accounts on X?}}
The number of followers and engagement metrics are used to inform us of the correlation between these metrics.
Table~\ref{tbl_corr_metrics} shows us that the followers metric has a strong correlation with likes, reposts, and replies but a weak correlation with shares. From all the engagement metrics, the shares metric has the lowest correlation when paired with the other engagement metrics. The two metrics with the highest and almost perfect correlation were the likes and reposts. 
This is expected because users who find content engaging, appealing, or insightful are highly likely to both like and share it. 

\input{tbl/combined_pvalue_table}

\subsection{\textbf{RQ3.1: Does the content of posts differ significantly across selected conferences?}}
The Chi-Square test along with the Bonferroni Correction for the content category scores from our manual classification of posts into their unique category indicated a significant difference with a p-value $<$0.05. 
For p-values $<$0.05, we identified that the NDSI (Networks: Computer Networks) conference had a high statistical significance in the category with all conferences as shown in Table~\ref{tbl_pvalues_combined}, the most out of any conference. The NSDI conference had the highest percentage of posts in the `Article and  Multimedia Sharing' category with 89\% of the posts.  The WSDM (Web \& Information Retrieval) and UIST (Human-Computer Interaction) conferences each had fifteen pairs of differences with other conferences. 
For the WSDM conference, 82\% of their posts were from the `Community Events' category. Similarly, UIST conference had 75\% posts from the `Community Events' category, which would explain the differences between these two conferences and the other selected conferences.

\subsection{\textbf{RQ3.2: Are there significant differences in sentiment scores across selected conferences?}}

A significant difference in conference sentiment scores is observed based on the Chi-Square test and subsequent Bonferroni Correction findings.
For p-values $<$0.05, we observed that NSDI and SIGMETRICS (Measurement \& Perf. Analysis) were tied with the most pairs of significance with other conferences with 12 as shown in Table~\ref{tbl_pvalues_combined}. IJCAI (Artificial Intelligence) followed with 11 pairs of significance with other conferences. 
We observed that 100\% of SIGMETRICS, 88\%  of NSDI, and  85\% of IJCAI conference posts were of neutral sentiment.

\subsection{\textbf{RQ3.3: Does the length of posts vary significantly across selected conferences?}}

The post's character length from our sample data of all conferences ranges from 16 to 350 and has a median of 194. The length of posts for the complete dataset of posts ranges from 0 to 1354, with a median value of 176.  
Fig.~\ref{fig-tweets-length} shows medians across conferences ranging from 68 to 268. It also shows that the medians across conferences for the complete dataset range from 79 to 249.

The findings from Kruskal-Wallis test, followed by Dunn Post-hoc test to both the sample and complete datasets, are presented in Table~\ref{tbl_pvalues_combined}. 
For p-value $<$0.05 we observed that in the sample dataset, SIGRAPH had the most pairs of significance with other conferences with 10, followed by WSDM with 7 and SIGMETRICS with 6 as shown in Table~\ref{tbl_pvalues_combined}. We also identified that ASE, ICDE, ICML, ICSA and NSDI all showed differences in sentiment with the same conferences: SIGGRAPH, SIGMETRICS and WSDM.
\input{fig/fig_rq3.3}
In our complete dataset, for p-values $<$0.05, we observed that EC, SIGGRAPH, SIGMETRICS and Stoc were the conferences that are significantly different in post length with 19 other conferences as shown Table~\ref{tbl_pvalues_combined}. ICML has a similar result,  having a significant difference in post length with 18 other conferences.
The EC (Economics \& Computation), SIGGRAPH (Computer Graphics), and SIGMETRICS (Measurement \& Perf. Analysis) had among the highest post length mean and median values, with SIGMETRICS having the highest mean post length followed by SIGGRAPH and EC conferences. The SIGGRAPH had a slightly higher median than the SIGMETRICS conference, but was then followed by the EC conference. The Stoc conference had one of the lowest post lengths in terms of both mean and median.

The analysis showed significant sentiment differences in the sample dataset and notable post-length variations in the complete dataset. 
These differences may stem from the nature of discussions, community engagement, or field-specific communication styles, with some conferences encouraging more detailed or technical exchanges than others.

\input{fig/figs_rq4}

\subsection{\textbf{RQ4: How do users on X engage with content posted by computer science conference accounts?}}

The most popular engagement metric used by X users to respond to content in both our sample dataset and complete dataset was the likes metric. The distribution of both datasets is displayed in Fig~\ref{fig-b}. The likes metric in the sample dataset across all conferences had a median of 9 and a mean of 21.94. The outlier in this dataset had a total of 269 likes, which was from Post ID 6153. Our complete dataset had similar ranges, with a median value of 7 and a mean of 28.27. This dataset had an outlier as well, with Post ID 1168 having a total of 27376 likes. 

The repost metric was the second most popular way users responded to content in both datasets. Fig~\ref{fig-d} displays the distribution of both datasets. Our sample dataset had a mean value of 4.63 and a median of 3, while our complete dataset had a mean of 7.79 and a median of 2. The sample dataset did not contain any outliers like our complete data, which had a tweet of 7513 reposts from Post ID 1168.  

The shares metric was the third most popular way users engage, followed by the replies metric. The shares metric in our sample dataset had a mean value of 0.67 and a median value of 0. Our complete dataset had a better mean value of 1.60 but had the same median value of 0. Fig.~\ref{fig-c} displays the data distribution for both datasets.

The replies metric was the least popular method the user interacted with in both datasets. The sample dataset had a mean value of 0.50 and a median value of 0.90 for our complete dataset. However, it had a median value of 0 for both datasets. Fig.~\ref{fig-a} contains the data distribution for both datasets.

A manual analysis of the posts from the sample dataset was performed to further understand the types of posts in the top 5 of each metric (Sec~\ref{sec_data})

We observed that Tweet ID 6153 had the highest number of likes and reposts and was fifth overall in the number of shares. However, it did not make the top-5 list for the tweet with the most replies but was one of two posts to appear in three top-5 lists, the other being Post ID 6147. Interestingly enough, Post ID 6129 which had the most replies did not appear in any of the other top-5 lists. Three posts appeared in two of the top-5 list of metrics: Post ID 9949 appeared in the shares and replies metric, Post ID 6130 appeared in the likes and reposts meitrcs, and Post ID 3267 appeared in the shares and reposts metrics. These posts were from the `Awards', `Others', and `Community Events' categories, respectively. 

We decided to further examine the content of the top posts. Tweet ID 6153 pertained to the `Awards' category referenced from Table~\ref{category_tbl}, this tweet congratulated the winners of the ICML Outstanding Paper Awards. We can deduce that this post had the most likes and reposts, since it reflects positive sentiment, which is why so many users liked and retweeted. Post ID 6129 belonged to the `Submissions' category, this post was asking its followers to post their post-paper submissions memories or celebration pictures. The post encourages users to post about their submission experience which is the reason why this particular post was the most commented. Post ID 3267 was the top post with the most shares, the content pertained to the `Community Events' category which posted the list of available CVPR workshops. Our observation as to why it is the leading post in the shares metric could be because users are reposting to make their followers aware while adding their comments to catch the attention of their followers.

Our manual analysis uncovered that the `Awards' category had the most total {posts} appearing in the top-5 lists. This could be because users want to express their excitement to posts that show recognition to those in the academic community.


%% file: tbl/tbl_conf_ratios.tex
\begin{table*}[htbp]

\caption{ Engagement and Influence Metrics of Our Selected Top Computer Science Conferences on X }

\label{tbl_conf_ratios}
\centering
\resizebox{\textwidth}{!}{%
\begin{tabular}{|l|l|p{0.09\linewidth}|l|l|l|l|l|l|l|l|}
\hline
   & X Handle  & Conference   & Category                         & Followers & Following & FFr                    & RMr (S)   & Ir (S)    & RMr (C) & Ir (C) \\ \hline
1  & icmlconf    & ICML    & Machine Learning                 & 74115     & 16        & \cellcolor[HTML]{8ED2B0}4632.2                 & 0.004 & 0.004 & 0.309    & 0.287   \\ \hline
2  & CVPR  &  CVPR          & Computer Vision                  & 44594     & 328       & 136                    & 0.005 & 0.004 & 0.209    & 0.178   \\ \hline
3  & siggraph & SIGGRAPH     & Computer Graphics                & 43666     & 1734      & 25.182                 & 0.001 & 0.001 & 0.396    & 0.364   \\ \hline
4  & usenix    & NSDI      & Networks: Computer Networks      & 14244     & 880       & 16.186                 & 0.002 & 0.002 & 0.165    & 0.135   \\ \hline
5  & emnlpmeeting  & EMNLP  & Natural Language Processing      & 13143     & 41        & 320.561                & 0.005 & 0.004 & 0.190    & 0.164   \\ \hline
6  & ieee\_ras\_icra & ICRA & Robotics                         & 10425        & 87     & 119.828                & 0.027 & 0.024 & 0.547    & 0.504   \\ \hline
7  & IEEESSP   &   Oakland    & Computer Security                & 9094      & 0         & \multicolumn{1}{c|}{-} & 0.022 & 0.020 & 0.810    & 0.744   \\ \hline
8  & IJCAIconf  &   IJCAI   & Artificial Intelligence          & 6776      & 831       & 8.154                  & 0.010 & 0.009 & 0.280    & 0.224   \\ \hline
9  & ACMUIST  & UIST     & Human-Computer Interaction       & 5302      & 67        & 79.134                 & 0.025 & 0.024 & \cellcolor[HTML]{8ED2B0}1.507    & \cellcolor[HTML]{8ED2B0}1.22    \\ \hline
10 & WSDMSocial &  WSDM    & The Web \& Information Retrieval & 3099      & 118       & 26.263                 & 0.035 & 0.029 & 0.982    & 0.921   \\ \hline
11 & ISCAConfOrg & ISCA     & Computer Architecture            & 3094      & 53        & 58.377                 & 0.004 & 0.004 & 0.250    & 0.225   \\ \hline
12 & IACReurocrypt & EuroCrypt   & CyberSecurity                    & 2601      & 3         & 867                    & 0.006 & 0.005 & 0.272    & 0.252   \\ \hline
13 & ASE\_conf  & ASE     & Software Engineering             & 1878      & 258       & 7.280                  & 0.040 & 0.036 & 1.112    & 0.938   \\ \hline
14 & AcmSIGecom  & EC    & Economic \& Computation          & 1833      & 3         & 611                    & 0.035 & 0.034 & 0.632    & 0.564   \\ \hline
15 & sigact & Stoc          & Algorithms \& Complexity         & 1548      & 4         & 387                    & 0.055 & 0.050 & 0.561    & 0.550   \\ \hline
16 & icdeconf   & ICDE     & Databases                        & 1052      & 0         & \multicolumn{1}{c|}{-} & 0.007 & 0.003 & 0.453    & 0.404   \\ \hline
17 & ACMMobiCom & MobiCom     & Mobile Computing                 & 908       & 43        & 21.116                 & 0.047 & 0.044 & 0.778    & 0.713   \\ \hline
18 & confCAV   & CAV      & Logic \& Verification            & 748       & 2         & 374                    & 0.032 & 0.032 & 0.758    & 0.650   \\ \hline
19 & sospconf  & SOSP      & Operating Systems                & 742       & 26        & 28.538                 & 0.020 & 0.020 & 0.460    & 0.416   \\ \hline
20 & ACMSigmetrics & SIGMETRICS   & Measurement \& Perf. Analysis    & 420       & 66        & 6.364                  & 0.017 & 0.017 & 0.479    & 0.457   \\ \hline
21 & PPoPPConf & PPoPP       & Programming Languages            & 246       & 38        & 6.474         & \cellcolor[HTML]{8ED2B0}0.057 & \cellcolor[HTML]{8ED2B0}0.057 & 0.720    & 0.634   \\ \hline
22 & RTASConf  & RTSS      & Embedded \& Real-Time Systems    & 206       & 27        & 7.630                  & 0.005 & 0.005 & 0.806    & 0.762   \\ \hline
\end{tabular}
}

{
\vspace{1mm}
Here, FFr: Follower/Following ratio, RMr: Repost and Mention ratio, Ir: Interactor ratio, (S): sample dataset, and (C): complete dataset.
}

\end{table*}

%% file: tbl/tbl_corr_conf.tex
\begin{table}[htb]
\caption{Correlation between X Engagement Metrics}

\label{tbl_corr_metrics}
\begin{tabular}{|l|l|l|l|l|l|}
\hline
          & Followers              & Replies               & Shares   & Reposts & Likes    \\ \hline
Followers & \multicolumn{1}{c|}{-} & 0.82               & 0.35 & 0.93 & 0.95 \\ \hline
Replies  & 0.82               & \multicolumn{1}{c|}{-} & 0.78 & 0.88 & 0.88 \\ \hline
Shares    & 0.35               & 0.78               & -        & 0.55 & 0.49 \\ \hline
Reposts     & 0.93               & 0.88               & 0.55 & -        & 0.98 \\ \hline
Likes  & 0.95               & 0.88               & 0.49 & 0.98 & -        \\ \hline
\end{tabular}
\end{table}

%% file: tbl/combined_pvalue_table.tex
\begin{table*}[htb]

\caption{
 Relationship Between Content Categories, Sentiment, and Post Length in Posts from Top Computer Science Conferences on X
}

\label{tbl_pvalues_combined}

\centering
\renewcommand{\arraystretch}{1.55}
\resizebox{\linewidth}{!}{
\begin{tabular}{lllllllllllllllllllllll}
           & \rotatebox{90}{ASE} & \rotatebox{90}{CAV} & \rotatebox{90}{CVPR} & \rotatebox{90}{EC } & \rotatebox{90}{EMNLP} & \rotatebox{90}{EuroCrypt} & \rotatebox{90}{ICDE} & \rotatebox{90}{ICML} & \rotatebox{90}{ICRA} & \rotatebox{90}{IJCAI} & \rotatebox{90}{ISCA} & \rotatebox{90}{MobiCom} & \rotatebox{90}{NSDI} & \rotatebox{90}{Oakland} & \rotatebox{90}{PPoPP} & \rotatebox{90}{RTSS} & \rotatebox{90}{ {\tiny \textbf{SIGGRAPH}}} & \rotatebox{90}{{\tiny \textbf{SIGMETRICS}}} & \rotatebox{90}{SOSP} & \rotatebox{90}{Stoc}  & \rotatebox{90}{UIST} & \rotatebox{90}{WSDM}\\
        
ASE        & - - - - & * - - * & * - - - & * - - * & * - - - & - - - *   & * - - * & - - - * & * - - - & * * - * & * - - - & - - - * & * * - * & * - - * & * - - - & * - - - & * - * *  & - * * *    & - - - - & * - - * & - - - - & - - * * \\
CAV        & * - - * & - - - - & - - - - & - - - * & * - - - & * - - *   & - - - * & - - - - & - - - - & * * - - & - - - * & * - - * & * * - - & - - - - & - - - * & - - - - & - - - *  & * * - *    & * - - - & * - - * & * - - * & * - - * \\
CVPR       & * - - - & - - - - & - - - - & - - - * & * - - - & - - - *   & - - - * & - - - * & - - - - & * * - * & * - - * & - - - * & * * - - & - - - * & - - - - & - - - - & * - * *  & - * - *    & * - - - & * - - * & * - - - & * - - * \\
EC         & * - - * & - - - * & - - - * & - - - - & - - - * & - - - *   & - - - * & - - - * & - * - * & * - - * & - - - * & - - - * & * - - * & - - - * & - - - * & - - - * & - - - -  & - - - -    & - - - * & - - - * & * * - * & * - - * \\
EMNLP      & * - - - & * - - - & * - - - & - - - * & - - - - & * - - *   & * - - * & - - - - & * - - - & * - - - & - - - * & * - - * & * * - - & * - - - & - - - * & - - - - & * - - *  & - * - *    & - - - - & - - - * & * - - * & * - - * \\
EuroCrypt  & - - - * & * - - * & - - - * & - - - * & * - - * & - - - -   & * - - - & - - - * & - - - * & * - - * & * - - - & - - - * & * - - * & - - - * & - - - - & - - - - & * - - *  & - - - *    & - - - * & * - - - & - - - * & - - - - \\
ICDE       & * - - * & - - - * & - - - * & - - - * & * - - * & * - - -   & - - - - & - * - * & - * - * & * - - * & * - - - & * - - - & * - - * & - * - * & - - - - & - - - - & - - * *  & * - * *    & * * - - & * - - * & * * - * & * * * - \\
ICML       & - - - * & - - - - & - - - * & - - - * & - - - - & - - - *   & - * - * & - - - - & - - - * & * * - * & * - - * & - - - * & * * - * & - - - - & - - - * & - - - * & * - * *  & - * * *    & - - - * & - - - * & * - - * & * - * * \\
ICRA       & * - - - & - - - - & - - - - & - * - * & * - - - & - - - *   & - * - * & - - - * & - - - - & * * - - & * - - * & - - - * & * * - - & - - - - & - - - * & - * - - & - - * *  & - * - *    & * - - - & * - - * & * - - * & * - * * \\
IJCAI      & * * - * & * * - - & * * - * & * - - * & * - - - & * - - *   & * - - * & * * - * & * * - - & - - - - & * - - * & * * - * & * - - - & * * - - & * - - * & - - - * & - * * *  & * - - *    & * * - - & * - - * & * * - * & * * - * \\
ISCA       & * - - - & - - - * & * - - * & - - - * & - - - * & * - - -   & * - - - & * - - * & * - - * & * - - * & - - - - & * - - - & * - - * & * - - * & - - - - & - - - - & * - * *  & * - * *    & - - - - & * - - * & * - - - & * - * - \\
MobiCom    & - - - * & * - - * & - - - * & - - - * & * - - * & - - - *   & * - - - & - - - * & - - - * & * * - * & * - - - & - - - - & * * - * & - - - * & - - - - & - - - - & * - * *  & - * - *    & * - - - & * - - * & - - - - & - - - * \\
NSDI       & * * - * & * * - - & * * - - & * - - * & * * - - & * - - *   & * - - * & * * - * & * * - - & * - - - & * - - * & * * - * & - - - - & * * - - & * - - * & * - - - & * * * *  & * - * *    & * * - - & * - - * & * * - * & * * * * \\
Oakland    & * - - * & - - - - & - - - * & - - - * & * - - - & - - - *   & - * - * & - - - - & - - - - & * * - - & * - - * & - - - * & * * - - & - - - - & - - - * & - - - * & - - - *  & * * - *    & * - - - & * - - * & * - - * & * - - * \\
PPoPP      & * - - - & - - - * & - - - - & - - - * & - - - * & - - - -   & - - - - & - - - * & - - - * & * - - * & - - - - & - - - - & * - - * & - - - * & - - - - & - - - - & - - - *  & - - - *    & - - - - & - - - - & * - - - & - - - - \\
RTSS       & * - - - & - - - - & - - - - & - - - * & - - - - & - - - -   & - - - - & - - - * & - * - - & - - - * & - - - - & - - - - & * - - - & - - - * & - - - - & - - - - & - - - *  & - - - *    & - - - - & - - - * & * - - - & * - - - \\
SIGGRAPH   & * - * * & - - - * & * - * * & - - - - & * - - * & * - - *   & - - * * & * - * * & - - * * & - * * * & * - * * & * - * * & * * * * & - - - * & - - - * & - - - * & - - - -  & * * - -    & - - - * & * - - * & * - * * & * - - * \\
SIGMETRICS & - * * * & * * - * & - * - * & - - - - & - * - * & - - - *   & * - * * & - * * * & - * - * & * - - * & * - * * & - * - * & * - * * & * * - * & - - - * & - - - * & * * - -  & - - - -    & - * - * & - - - * & - * * * & - * - * \\
SOSP       & - - - - & * - - - & * - - - & - - - * & - - - - & - - - *   & * * - - & - - - * & * - - - & * * - - & - - - - & * - - - & * * - - & * - - - & - - - - & - - - - & - - - *  & - * - *    & - - - - & - - - * & - - - - & * - - * \\
Stoc       & * - - * & * - - * & * - - * & - - - * & - - - * & * - - -   & * - - * & - - - * & * - - * & * - - * & * - - * & * - - * & * - - * & * - - * & - - - - & - - - * & * - - *  & - - - *    & - - - * & - - - - & * - - * & * - - * \\
UIST       & - - - - & * - - * & * - - - & * * - * & * - - * & - - - *   & * * - * & * - - * & * - - * & * * - * & * - - - & - - - - & * * - * & * - - * & * - - - & * - - - & * - * *  & - * * *    & - - - - & * - - * & - - - - & - - * * \\
WSDM       & - - * * & * - - * & * - - * & * - - * & * - - * & - - - -   & * * * - & * - * * & * - * * & * * - * & * - * - & - - - * & * * * * & * - - * & - - - - & * - - - & * - - *  & - * - *    & * - - * & * - - * & - - * * & - - - -

\end{tabular}
}
\vspace{2mm}
{
Here, 
Each cell comprises [$x_1 x_2 x_3 x_4$] 
where 
$x_1$: X-post category,
$x_2$: X-post sentiment,
$x_3$: X-post length (Sampled), and 
$x_4$: X-post length (Complete);

Also, $x \in $ \{-,$\ast$\} where $\ast$: p-value$<$0.05, -:  p-value$\geq$0.05 .

}
\end{table*}

%% file: fig/fig_rq3.3.tex
\begin{figure}
    \centering
    \includegraphics[width=\columnwidth]{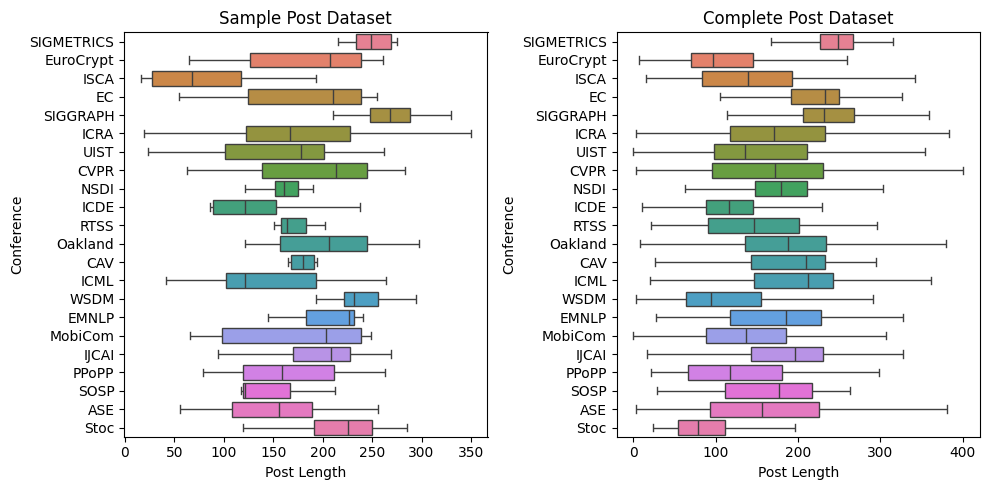}
    \caption{Conferences' X-Post Length}
    \label{fig-tweets-length}
\end{figure}

%% file: fig/figs_rq4.tex
\begin{figure}
    \centering
    \includegraphics[width=\columnwidth]{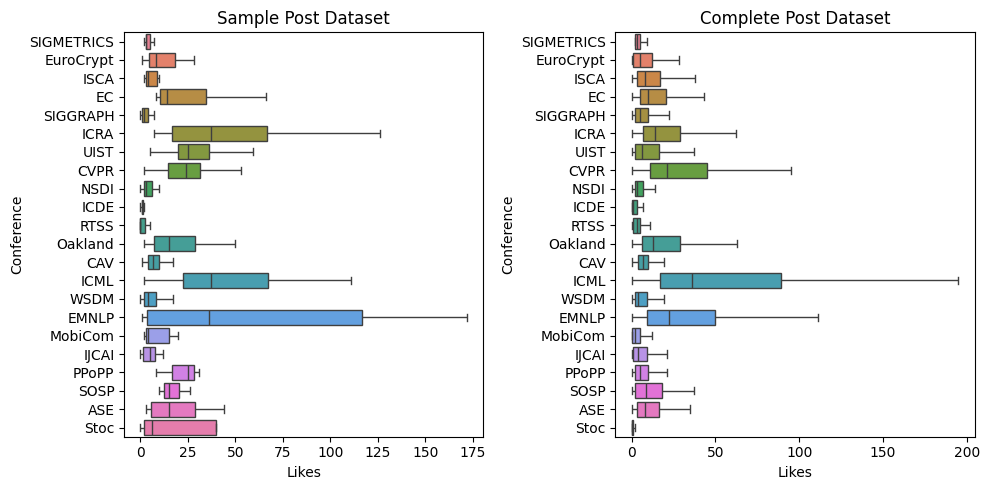}
    \caption{User Likes on Conferences' Posts on X}
    \label{fig-b}
\end{figure}

\begin{figure}
    \centering
    \includegraphics[width=\columnwidth]{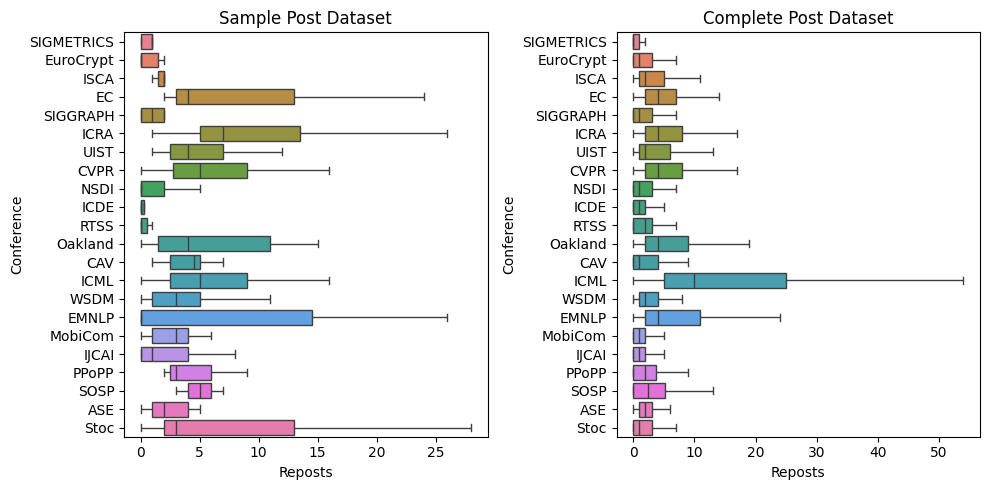}
    \caption{User Reposts on Conferences' Posts on X}
    \label{fig-d}
\end{figure}

\begin{figure}
    \centering
    \includegraphics[width=\columnwidth]{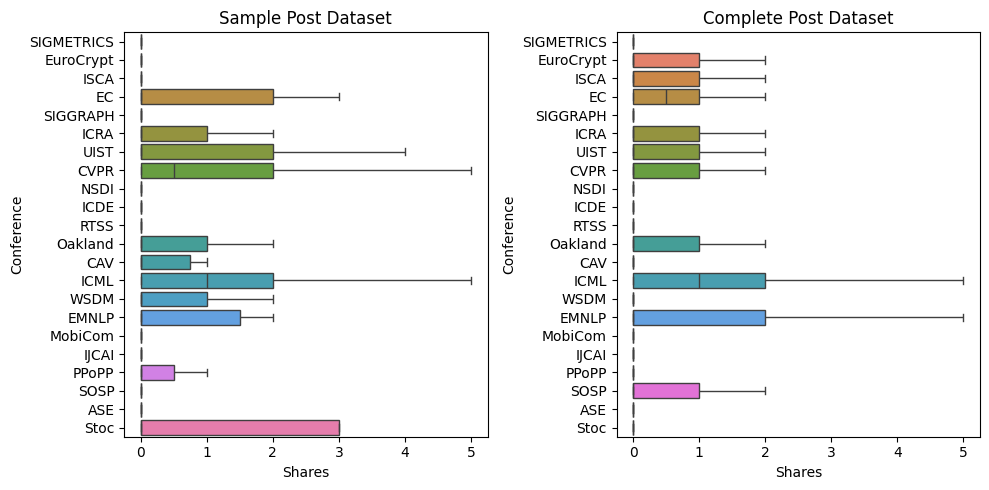}
    \caption{User Shares on Conferences' Posts on X}
    \label{fig-c}
\end{figure}

\begin{figure}
    \centering
    \includegraphics[width=\columnwidth]{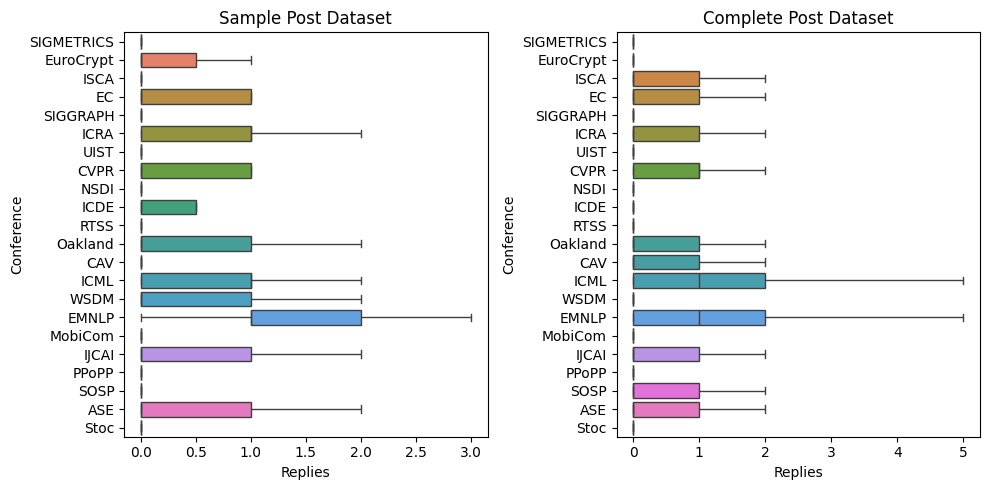}
    \caption{User Replies on Conferences' Posts on X}
    \label{fig-a}
\end{figure}

%% file: sec_threats.tex
In this section, we outline potential threats to the validity of our study design.  

\textbf{External validity} concerns the generalizability of our findings beyond the studied dataset. Our analysis focused on 22 computer science conference accounts on X, spanning various subfields. While our results provide insights into engagement patterns within computer science conferences, they may not be applicable to conferences in other disciplines. However, further studies would be needed to confirm the broader applicability of our findings across different academic communities.  

\textbf{Construct validity} ensures that our study accurately measures the theoretical concepts under investigation. In our manual annotation process, we adopted methodologies from previous studies to ensure reliability. To minimize subjectivity and bias, two independent annotators manually classified the tweets, discussing and resolving any disagreements until they reached full consensus. This approach strengthens the validity of our classification framework.  

\textbf{Internal validity} assesses whether the study effectively answers the research questions while minimizing bias and errors. A potential threat in our study lies in the accuracy of our sample dataset, given the large volume of posts. To mitigate this risk, we applied a 95\% confidence interval to ensure a robust and representative sample for manual analysis. This statistical approach helped maintain the integrity of our dataset while balancing feasibility and accuracy.

%% file: sec_conclusion.tex
X (formerly Twitter) plays a significant role in the academic community, serving as a backchannel for conference attendees and facilitating discussions beyond physical events. While previous studies have analyzed how attendees engage with social media during conferences, our study shifts the focus to the official conference accounts of 22 computer science conferences over a 14-year period. Our goal was to explore the relationship between engagement metrics and follower count, as well as to investigate statistical differences in tweet length, sentiment, and content categories across conferences.  

Our findings reveal that conferences tailor their posts differently based on their target audience, and statistical variations exist in their posting patterns, sentiment, and content themes. Additionally, we found a correlation between follower count and engagement metrics, providing insights into how visibility and interaction levels vary across conferences. Notably, likes emerged as the most common form of engagement with conference posts.  

While prior research has primarily focused on attendee-generated content, our study highlights the need for further investigation into how conferences themselves leverage social media for engagement. Future research could expand beyond X to examine other platforms, such as LinkedIn, Mastodon, or Bluesky, to gain a more comprehensive understanding of how academic conferences interact with their communities in the evolving social media landscape.